\pdfoutput=1 %for arxiv only
\documentclass[english,prd,floatfix,superscriptaddress,nofootinbib]{revtex4-2}
\usepackage[T1]{fontenc}
\usepackage[utf8]{inputenc}
\usepackage{color}
\definecolor{note_fontcolor}{rgb}{0.800781, 0.800781, 0.800781}
\usepackage{babel}
\usepackage{mathtools}
\usepackage{amsmath}
\usepackage{amssymb}
\usepackage{graphicx}
\usepackage{rotfloat}
\PassOptionsToPackage{hyphens}{url}\usepackage[unicode=true,pdfusetitle,
bookmarks=true,bookmarksnumbered=false,bookmarksopen=false,
breaklinks=false,pdfborder={0 0 0},pdfborderstyle={},backref=false,colorlinks=false]{hyperref}

\makeatletter

\usepackage{colortbl}
\usepackage{upgreek}
\usepackage{url}

\usepackage{tensind}
\tensordelimiter{?}
\usepackage{pgfplots}
\usetikzlibrary{pgfplots.groupplots}

\DeclareSymbolFont{vectors}{OML}{cmm}{b}{it}
\DeclareSymbolFont{tensors}{OT1}{cmss}{bx}{it}

\DeclareSymbolFontAlphabet{\mathvec}{vectors}
\DeclareSymbolFontAlphabet{\mathtens}{tensors}
\DeclareUnicodeCharacter{03B2}{\ensuremath{\upbeta}}

\usepackage[capitalise]{cleveref}

\makeatother

\begin{document}
\global\long\def\tudu#1#2#3#4{{?{#1}^{#2}{}_{#3}{}^{#4}?}}%

\global\long\def\tud#1#2#3{{?{#1}^{#2}{}_{#3}?}}%

\global\long\def\tudud#1#2#3#4#5{{?{#1}^{#2}{}_{#3}{}^{#4}{}_{#5}?}}%

\global\long\def\tdu#1#2#3{\tensor{#1}{_{#2}^{#3}}}%

\global\long\def\dd#1#2{\frac{\mathrm{d}#1}{\mathrm{d}#2}}%

\global\long\def\pd#1#2{\frac{\partial#1}{\partial#2}}%

\global\long\def\tens#1{\mathtens{#1}}%

\global\long\def\threevec#1{\mathvec{#1}}%

\global\long\def\d{\mathrm{d}}%

\global\long\def\e{\mathrm{e}}%

\global\long\def\eps{\varepsilon}%

\global\long\def\i{\mathrm{i}}%

\global\long\def\ext{\tilde{\mathrm{d}}}%

\title{Self-interactions can (also) destabilize bosonic stars}
\author{Marco Brito}
\email[]{marcobrito@ua.pt}
\affiliation{Departamento de Matemática da Universidade de Aveiro \\
and Centre for Research and Development in Mathematics and Applications (CIDMA) \\
Campus de Santiago, 3810-193 Aveiro, Portugal}
\author{Carlos Herdeiro}
\affiliation{Departamento de Matemática da Universidade de Aveiro \\
and Centre for Research and Development in Mathematics and Applications (CIDMA) \\
Campus de Santiago, 3810-193 Aveiro, Portugal}
\author{Nicolas Sanchis-Gual}
\affiliation{Departamento de Astronomía y Astrofísica, Universitat de València, \\
Dr.\ Moliner 50, 46100, Burjassot (Valencia), Spain}
\author{Etevaldo dos Santos Costa Filho}
\affiliation{Departamento de Matemática da Universidade de Aveiro \\
and Centre for Research and Development in Mathematics and Applications (CIDMA) \\
Campus de Santiago, 3810-193 Aveiro, Portugal}
\author{Miguel Zilhão}
\affiliation{Departamento de Matemática da Universidade de Aveiro \\
and Centre for Research and Development in Mathematics and Applications (CIDMA) \\
Campus de Santiago, 3810-193 Aveiro, Portugal}

\begin{abstract}
We study the dynamical stability of Proca-Higgs stars, in spherical symmetry. These are solutions of the Einstein-Proca-Higgs model, which features a Higgs-like field coupled to a Proca field, both of which minimally coupled to the gravitational field. The corresponding stars can be regarded as Proca stars with
self-interactions, while avoiding the hyperbolicity issues of self-interacting Einstein-Proca models.
We report that these configurations are stable near the Proca limit
in the candidate stable branches, but exhibit instabilities in certain
parts of the parameter space, even in the candidate stable branches, regaining
their stability for very strong self-interactions. This shows that for these models, unlike various examples of \textit{scalar} boson stars, 
self-interactions can deteriorate, rather than improve, the dynamical robustness of bosonic stars. 
\end{abstract}
\maketitle

\section{Introduction}

Bosonic stars are localized, horizonless, self-gravitating lumps of bosonic matter,
sustained by their self-gravity and possibly additional self-interaction
forces, see Refs.~\citep{Liebling:2012fv,Schunck:2003kk} for reviews.
These hypothetical astrophysical objects are an alternative to (or
coexisting with) astrophysical black holes \citep{Schunck:1998cdq,Mielke:2000mh,Berti:2006qt,Guzman:2009zz,Herdeiro:2022yle}. Amongst various interesting phenomenological features, their dynamics can match real gravitational wave
signals during the merger of these objects \citep{CalderonBustillo:2020fyi,CalderonBustillo:2022cja} as well as the imaging properties of black holes, under special circumstances~\cite{Vincent:2015xta,Cunha:2015yba,Cunha:2016bjh,Cunha:2017wao,Grould:2017rzz,Olivares:2018abq,Herdeiro:2021lwl,Rosa:2022tfv,Sengo:2022jif,Rosa:2022toh,Rosa:2023qcv,Rosa:2024eva,Sengo:2024pwk}.

Several models of bosonic stars have been constructed: from scalar boson stars, e.g.~\citep{Kaup:1968zz,Ruffini:1969qy,Colpi:1986ye,Bernal:2009zy,Hartmann:2013tca,Herdeiro:2017fhv,Alcubierre:2018ahf,Brihaye:2018grv,Guerra:2019srj,Boskovic:2021nfs,Maso-Ferrando:2021ngp,Sanchis-Gual:2021phr,Brito:2023fwr},
which are described (in the simplest models) by a complex scalar field which is  minimally coupled to the gravitational
field, to vector stars, also known as Proca stars \citep{Brito:2015pxa} -- see also e.g.~\cite{SalazarLandea:2016bys,Duarte:2016lig,Minamitsuji:2017pdr,Herdeiro:2019mbz,Herdeiro:2020jzx,Dzhunushaliev:2021vwn,Brihaye:2021qvc,Aoki:2022mdn,Aoki:2022woy,Ma:2023vfa,Jockel:2023rrm,Pombo:2023sih,Hernandez:2023tig,Wang:2023tly,Ma:2023bhb,Zhang:2023rwc,Lazarte:2024jyr}. These models are composed (in the simplest models) by complex a vector
field minimally  coupled to gravity.
Although scalar and vector stars have similarities \citep{Herdeiro:2017fhv,Herdeiro:2019mbz,Sanchis-Gual:2018oui},
they differ in some key aspects, such as their geodesic structure (see e.g.~\cite{Delgado:2021jxd}), which may (or not) have features leading to the ability to mimic the shadow of black
holes \citep{Herdeiro:2021lwl,Sengo:2024pwk}, and their stability. Indeed, static, spherical Proca stars are unstable against non-spherical dynamics, unlike scalar mini-boson stars, having as their
ground state an axisymmetric configuration~\citep{Herdeiro:2023wqf}. Curiously, in the simplest models, this (in)stability is reversed in the case of stationary (rotating) stars, with the mini-Proca stars being stable and the scalar stars being unstable (against a non-axisymmetric mode)~\cite{Sanchis-Gual:2019ljs}. 

Another major difference between scalar and vector fields within these classes of models pertains self-interactions. While for scalar models self-interactions are theoretically consistent and might increase the stability of the stars in these models~\cite{Siemonsen:2020hcg,Sanchis-Gual:2021phr,Ildefonso:2023qty}, for the Proca models, despite being theoretically justified across various phenomenological scenarios, self-interactions introduce significant issues, varying from the loss of hyperbolicity of the field equations to new sources of instabilities \cite{Coates:2022nif,Clough:2022ygm,Mou:2022hqb,Coates:2022qia,Barausse:2022rvg,Hell:2024xbv}. This has led to serious concerns of Einstein-Proca models where the Proca field has self-interactions.

An apparently unrelated concern is that the vector (or Proca) stars require a model with a mass term, which breaks
gauge symmetry and is typically postulated ``ad hoc'' in the action. It could be expected that such mass is actually generated by a Higgs mechanism, similar to the one for
vector bosons in the Standard Model. This has motivated coupling the vector field to a real Higgs-like scalar field \citep{Herdeiro:2023lze}. As it turns out,  this formulation effectively endows the vector field not only with mass, but also with self-interactions which, moreover, avoids the hyperbolicity problems in the (self-interacting) Einstein-Proca theory \citep{Barausse:2022rvg} and can be viewed as a UV completion of the self-interacting Einstein-Proca models. The Einstein-Proca-Higgs model yields solitonic solutions known as Proca-Higgs stars, with the usual mini-Proca stars arising for certain limiting values
of the Higgs boson mass and magnitude of the Higgs potential.

Solutions to the Einstein-Proca-Higgs system have already been obtained
before \citep{Herdeiro:2023lze}, but to ascertain these stars' physical
relevance, it is needed to study their dynamical robustness. The minimal (and technically simplest) test is to study radial dynamics. This analysis is the goal of our paper, for which we will perform 1+1D numerical relativity simulations of the spherical Proca-Higgs stars.

Since the Einstein-Proca-Higgs model provides a way to include self-interactions in a consistent manner in a Proca model, avoiding
the problems associated with (pure) self-interacting Proca models, such as the loss of hyperbolicity, we are able to successfully
evolve Proca-Higgs stars and test the impact of the self-interactions. These stars reduce to mini-Proca stars in the limit
where self-interactions vanish, so we may inquire if increasing the
interactions keeps the stars stable (within spherical dynamics) or induces an instability. Such a case would
contrast with scalar boson stars, where self-interactions can heal the inherent
instabilities of excited configurations.
We will evolve static models, first near the Proca limit to test the
robustness of our models, which should agree with the mini-Proca ones,
and then explore the parameter space further away from the Proca limit to test the stability therein. As we shall see, our results show the self-interactions can deteriorate the stability of the spherical Proca-Higgs stars in regions of the parameter space where, naively, stability could be expected. Thus, our study 
complements the picture on the impact of self-interactions on the dynamics of bosonic stars, highlighting a different guise of the contrast between scalar and vector models.

This paper is organized as follows. In Section~\ref{sec2} the Einstein-Proca-Higgs model is detailed, together with a brief discussion of the equilibrium solutions describing spherical Proca-Higgs stars. In Section~\ref{sec3} the framework for the numerical evolutions is presented. The numerical evolutions are presented in Section~\ref{sec4} and a discussion of the results together with conclusions are presented in Section~\ref{sec5}. An appendix discusses numerical convergence.

\section{The Einstein-Proca-Higgs action}
\label{sec2}

\subsection{Action and equations of motion}

The Einstein-Proca-Higgs model \citep{Herdeiro:2023lze} is derived from the following action
\begin{equation}
S[g_{\mu\nu},\phi,A_{\mu}]=\int\d^{4}x\sqrt{-g}\left[\frac{R}{16\pi G}-\frac{1}{4}F_{\mu\nu}\bar{F}^{\mu\nu}-\frac{1}{2}\phi^{2}A_{\mu}\bar{A}^{\mu}-\frac{1}{2}\partial_{\mu}\phi\partial^{\mu}\phi-\frac{\lambda}{4}(\phi^{2}-v^{2})^{2}\right] \ ,
\end{equation}
where $v$ is the vacuum expectation value (VEV), $\lambda$ is the
self-interaction constant, $F_{\mu\nu}=\partial_{\mu}A_{\nu}-\partial_{\nu}A_{\mu}$
and we use $c=\hbar=1$. The field $A$ denotes a complex vector field, while $\phi$ denotes a real scalar field. The dimensions of the fields are $[\phi]=[A_{t(r)}]=L^{-1}$.
This results in the equations for the fields
\begin{align}
\nabla_{\mu}F^{\mu\nu} & =\phi^{2}A^{\nu} \ , \\
\square\phi&=\lambda\phi(\phi^{2}-v^{2})+\phi A_{\mu}\bar{A}^{\mu} \ ,
\end{align}
with the requirement that the ``Lorenz gauge condition'' be satisfied
\begin{equation}
\nabla_{\mu}(\phi^{2}A^{\mu})=0\ ,
\end{equation}
which, unlike Maxwell's theory, is a dynamical requirement. This
also happens in the Proca case. We also get the Einstein equations
\begin{equation}
R_{\mu\nu}-\frac{1}{2}Rg_{\mu\nu}=8\pi G\left[T_{\mu\nu}^{(v)}+T_{\mu\nu}^{(s)}\right] \ ,
\end{equation}
where the vector and scalar components of the stress-energy tensor
are
\begin{equation}
T_{\mu\nu}^{(v)}\coloneqq\frac{1}{2}(F_{\mu\sigma}\bar{F}_{\nu\gamma}+\bar{F}_{\mu\sigma}F_{\nu\gamma})g^{\sigma\gamma}-\frac{1}{4}g_{\mu\nu}F_{\sigma\tau}\bar{F}^{\sigma\tau}+\phi^{2}\left[\frac{1}{2}(A_{\mu}\bar{A}_{\nu}+\bar{A}_{\mu}A_{\nu})-\frac{1}{2}g_{\mu\nu}A_{\sigma}\bar{A}^{\sigma}\right] \ ,
\end{equation}
and
\begin{equation}
T_{\mu\nu}^{(s)}\coloneqq\partial_{\mu}\phi\partial_{\nu}\phi-g_{\mu\nu}\left[\frac{1}{2}\partial_{\rho}\phi\partial^{\rho}\phi+U(\phi)\right] \ ,
\end{equation}
where $U(\phi)=\frac{\lambda}{4}(\phi^{2}-v^{2})^{2}$.
We work with dimensionless quantities \citep{Herdeiro:2023lze}, scaling
the fields as
\[
A'_{\mu}\coloneqq A_{\mu}/v, \quad
\phi'\coloneqq \phi/v, \quad 
r'\coloneqq rv, \quad
\omega'\coloneqq \omega/v \ ,
\]
where $\omega$ is the frequency of the field.
In what follows, we will use as parameters $\lambda$ and $\alpha\coloneqq\sqrt{4\pi G}v$.
The field equations will take the form (where we drop the primes)
\begin{align}
& \nabla_{\mu}F^{\mu\nu}=\phi^{2}A^{\nu} \ , \\
& \square\phi=\lambda(\phi^{2}-1)\phi+\phi A_{\mu}\bar{A}^{\mu} \ , \\
& R_{\mu\nu}-\frac{R}{2} g_{\mu\nu}=2\alpha^{2}\left[T_{\mu\nu}^{(v)}+T_{\mu\nu}^{(s)}\right]\ .\label{eq:new_einst}
\end{align}

\subsection{Ansatz, boundary conditions and solutions}

We are analyzing the Einstein-Proca-Higgs system assuming spherical symmetry;
so in order to solve Einstein's equations we can consider a metric ansatz given in isotropic coordinates to be
of the form
\begin{equation}
\d s^{2}=-\e^{2F_{0}(r)}\d t^{2}+\e^{2F_{1}(r)}[\d r^{2}+r^{2}(\d\theta^{2}+\sin^{2}\theta\,\d\varphi^{2})] \ ,
\end{equation}
and for the fields
\begin{align}
\tilde{A}&=\e^{-\i\omega t}[f(r)\d t+\i g(r)\d r] \ ,\label{eq:proca-decomp-initial} \\
\phi&=\phi(r) \ .
\end{align}

To solve the field equation and obtain Proca-Higgs stars, we impose asymptotic flatness which requires that
\begin{align*}
F_{0,1}(r\to\infty) & =0 \ ,\\
\phi(r\to\infty) & =v \ ,\\
f(r\to\infty)=g(r\to\infty) & =0 \ .
\end{align*}
Additionally, we impose regularity at the origin, which implies
\[
\partial_{r}F_{0,1}(0)=\partial_{r}f(0)=\partial_{r}g(0)=\partial_{r}\phi(0)=0 \ .
\]

\begin{figure*}[tpbh]
\begin{centering}
\input{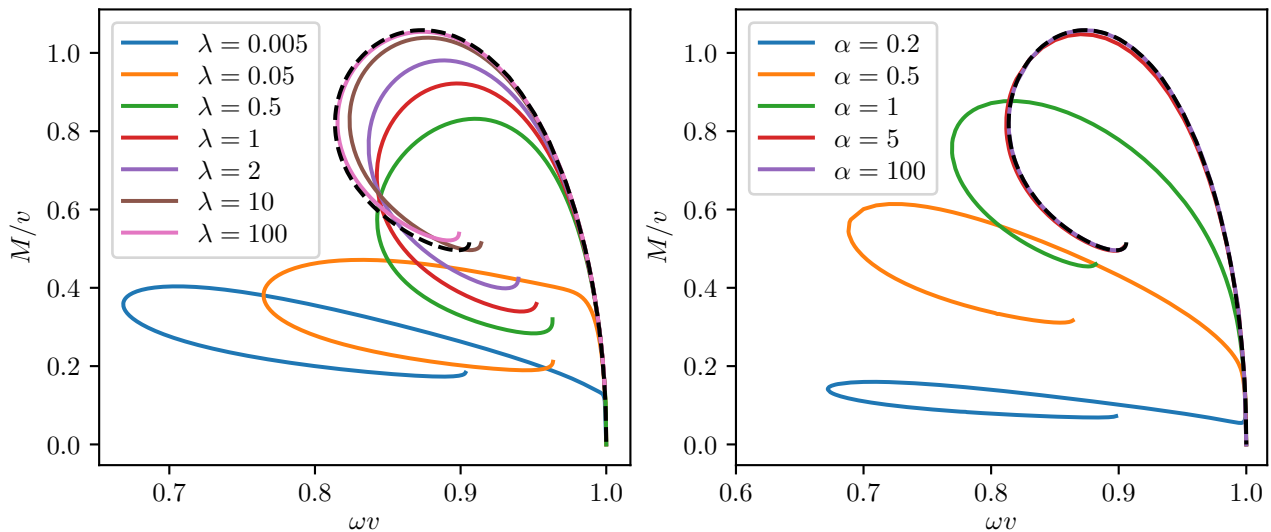}
\end{centering}
\caption{\label{fig:Mass-vs-frequency}Mass vs frequency diagram for Proca-Higgs
boson stars, with $\alpha=0.35$ of the left and $\lambda=0.005$  on the right. The black dotted line denotes mini-Proca stars.}
\end{figure*}

We solve the system above to obtain
a continuous family of solutions of Proca-Higgs stars for
each $\lambda$ and $\alpha$, parameterized by $\omega$ (with a discrete degeneracy), obeying
the condition $\omega_{{\rm min}}\leq\omega\leq v$. This is done by finding the solutions to the coupled non-linear differential equations for the functions $\mathcal{F}=(F_0 ,F_1 ;\phi ,f,g)$ using the professional package \texttt{FIDISOL/CADSOL} \cite{Schonauer}. More information can be found in \cite{Herdeiro:2023lze}. The solutions
are located as usual on spiral-type curves relating the Arnowitt-Deser-Misner
(ADM) mass with the frequency $\omega$ of the stars. Some examples can be found
in \cref{fig:Mass-vs-frequency}.
We divide the curve of solutions into a candidate stable branch -- ranging from the maximal frequency until the mass attains a maximum -- where according to turning point arguments \cite{Santos:2024vdm}
stars are stable under infinitesimal perturbations that conserve the
total mass and particle number, and the unstable branch where the
solutions are perturbatively unstable, and thus
physically irrelevant. Our goal is to test whether the Proca-Higgs stars are indeed dynamically stable throughout the parameter space in the candidate stable branch.

\subsection{Effective theory}
As mentioned before, one can consider this model as a UV completion of the Einstein-Proca model \cite{Herdeiro:2023lze,Aoki:2022woy}.
Expanding the scalar field around the VEV one obtains the effective Lagrangian
\begin{equation}
  \mathcal{L}^{\text{Proca EFT}}= - \frac{1}{4}F_{\rho\sigma}\bar{F}^{\rho\sigma}-\frac{\mu^2}{2}A_\rho \bar{A}^\rho
  -\alpha_2 (A_\rho \bar{A}^\rho)^2
\end{equation}
where $\mu=v$ and $\alpha_2=-1/4\lambda$ and this determines the strength of the self-interactions.
We see that the self-interactions are suppressed when $\alpha$ or $\lambda$ are very large and in fact 
we are left with just a mini-Proca star. To explore the parameter space, we will start from Proca-Higgs stars
with negligible self-interactions -- i.e.\ essentially mini-Proca Stars -- to stars with strong interactions, possibly outside the regime of
validity of the effective theory.

\section{Numerical evolutions -- framework}
\label{sec3}

\subsection{BSSN formalism and basic equations}

Using a standard 3+1 spacetime decomposition~\cite{Gourgoulhon:2007ue}, we can write the spacetime
metric as
\begin{equation}
\d s^{2}=-N^{2}\d t^{2}+h_{ij}(\d x^{i}+\beta^{i}\d t)(\d x^{i}+\beta^{j}\d t) \ ,
\end{equation}
where, as usual, $N$ denotes the lapse function, $\beta^{i}$ the shift
functions and $h_{ij}$ the induced metric of the spatial three-dimensional hypersurfaces.
Our induced metric can be written as
\[
\d l^{2}=\e^{4\chi}[a(t,r)\d r^{2}+r^{2}b(t,r)\d\Omega^{2}] \ ,
\]
where $\d\Omega^{2}=\d\theta^{2}+\sin^{2}\theta\,\d\varphi^{2}$ and
$a(t,r)$ and $b(t,r)$ are two non-vanishing conformal metric functions,
being related to the physical metric by the conformal decomposition
$h_{ij}=\e^{4\chi}\hat{h}_{ij}$ with $\e^{\chi}=(h/\hat{h})^{1/12}$,
where $h$ and $\hat{h}$ are the determinant of the physical
and conformal 3-metrics, respectively. We use the Baumgarte-Shapiro-Shibata-Nakamura
(BSSN) formulation \citep{Baumgarte:1998te,Shibata:1995we} in spherical
coordinates \citep{Alcubierre:2011pkc,Montero:2012yr,SanchisGual:2014ewa},
which are suited for the problem at hand. We also split the Proca field into its scalar $A_\Phi$ and vector $A_i$ potentials defined as
\begin{equation}
    \prescript{(3)}{}{A}_i = h^{\mu}_i A_\mu,\text{and}\quad A_{\Phi}=-n^{\mu}A_\mu,
\end{equation}
where $h^{\mu}_{\nu}$ is the projector operator onto the spatial hypersurfaces and $n^{\mu}$ is the timelike unit vector normal to the hypersurfaces.

The equations of motion can be decomposed as well into 6 field variables,
$E^{r},A_{r},A_{\Phi},\Gamma,\phi,K_{\phi}$, where
\[E^\mu = -n_\nu F^{\nu \mu}\ ,
\]
\[K_\phi = -\frac{1}{2\alpha}(\partial_t -\beta^i \partial_i)\phi \ ,\] $\Gamma=D_i \prescript{(3)}{}{A}^i$,  and assuming spherical symmetry, and we get for
the vector field
\begin{align}
(\partial_{t}-\pounds_{\vec{\beta}})E^{r}&=NKE^{r}+N\phi^{2}h^{rr}\prescript{(3)}{}{A}_{r} \ ,\\
(\partial_{t}-\pounds_{\vec{\beta}})\prescript{(3)}{}{A}_{r}&=-NE_{r}- A_{\Phi}\partial_{r}N-N\partial_{r}A_{\Phi} \ ,\\
\phi^{2}(\partial_{t}-\pounds_{\vec{\beta}}) A_{\Phi}&=2 A_{\Phi}\phi(-2NK_{\phi})+\phi^2 (A_{\Phi}K+N\Gamma+h^{rr}\prescript{(3)}{}{A}_{r}\partial_{r}N-N)+2Nh^{rr}\prescript{(3)}{}{A}_{r}\phi\partial_{r}\phi\label{eq:one-over-sphi} \ ,\\
\partial_{t}\Gamma&=-E^{r}\partial_{r}N+N\phi^{2} A_{\Phi}-ND^{i}D_{i} A_{\Phi}-2h^{rr}\partial_{r} A_{\Phi}\partial_{r}N- A_{\Phi}D^{i}D_{i}N+D^{i}(\pounds_{\vec{\beta}}\prescript{(3)}{}{A}_{i}) \ ,
\end{align}
where $\pounds_{\vec{\beta}}$ is the Lie derivative with respect to the shift vector $\vec{\beta}$ and $D_{i}$ is the covariant derivative associated with the spatial 3-metric $h_{ij}$. For the scalar field we get
\begin{align}
(\partial_{t}-\pounds_{\vec{\beta}})\phi &=-2NK_{\phi} \ , \\
(\partial_{t}-\pounds_{\vec{\beta}})K_{\phi}&=NKK_{\phi}+\frac{N}{2}\left[\lambda(\phi^{2}-v^{2})\phi+\phi(-|A_{\Phi}|^{2}+h^{rr}\prescript{(3)}{}{A}_{r}\prescript{(3)}{}{\bar{A}}_{r})\right]-\frac{1}{2}h^{rr}\phi_{,r}\partial_{r}N-\frac{1}{2}h^{ij}ND_{j}\phi_{,i} \ ,
\end{align}
where
\begin{equation}
h^{ij}D_{j}\phi_{,i}=\frac{1}{a\e^{4\varphi}}\left[\partial_{r}\psi-\psi\left(\frac{\partial_{r}a}{2a}-\frac{\partial_{r}b}{b}-\frac{2}{r}-2\partial_{r}\varphi\right)\right]\ .
\end{equation}
As for the components of the stress-energy tensor we define the energy density and the spatial part of the tensor as $\rho=n^\alpha n^\beta T_{\alpha \beta}$, $S_a=S^r_r$ and $S_b=S^\theta_\theta$, where $S_{ij}=\tud h{\alpha}{i}\tud h{\beta}{j}T_{\alpha\beta}$. Then
\begin{align*}
\rho^{(s)}&=\frac{1}{2N^{2}}(\partial_{t}\phi)^{2}-\frac{1}{N^{2}}\beta^{r}\partial_{t}\phi\partial_{r}\phi+\frac{\beta^{r}\beta^{r}}{2N^{2}}\partial_{r}\phi\partial_{r}\phi+\frac{1}{2}h^{rr}\partial_{r}\phi\partial_{r}\phi+\frac{\lambda}{4}(\phi^{2}-v^{2})^{2} \ , \\
j^{(s)r}&=-\frac{1}{N}h^{rr}\partial_{r}\phi(\partial_{t}\phi-\beta^{r}\partial_{r}\phi) \ , \\
S_{a}^{(s)}&=\frac{1}{2}h^{rr}\partial_{r}\phi\partial_{r}\phi-\left(-\frac{1}{2N^{2}}(\partial_{t}\phi)^{2}+\frac{1}{N^{2}}\beta^{r}\partial_{t}\phi\partial_{r}\phi-\frac{\beta^{r}\beta^{r}}{2N^{2}}\partial_{r}\phi\partial_{r}\phi+\frac{\lambda}{4}(\phi^{2}-v^{2})^{2}\right) \ , \\
S_{b}^{(s)}&=\frac{1}{2N^{2}}(\partial_{t}\phi)^{2}-\frac{1}{N^{2}}\beta^{r}\partial_{t}\phi\partial_{r}\phi+\frac{\beta^{r}\beta^{r}}{2N^{2}}\partial_{r}\phi\partial_{r}\phi-\frac{1}{2}h^{rr}\partial_{r}\phi\partial_{r}\phi-\frac{\lambda}{4}(\phi^{2}-v^{2})^{2} \ ,
\end{align*}
and for the vector part
\begin{align*}
\rho^{(v)}&=\frac{1}{2}h_{rr}E^{r}\bar{E}^{r}+\frac{1}{2}\phi^{2}\left[{ A}_{\Phi}\bar{{ A}}_{\Phi}+h^{rr}\prescript{(3)}{}{A}_{r}\prescript{(3)}{}{\bar{A}}_{r}\right] \ , \\
j^{(v)r}&=\frac{1}{2}\phi^{2}h^{rr}(\prescript{(3)}{}{A}_{r}\bar{{ A}}_{\Phi}+\prescript{(3)}{}{\bar{A}}_{r}{ A}_{\Phi}) \ , \\
S_{a}^{(v)} &=-\frac{1}{2}E^{r}\bar{E}_{r}+\phi^{2}\left(\frac{1}{2}\prescript{(3)}{}{A}^{r}\prescript{(3)}{}{\bar{A}}_{r}+\frac{1}{2}|{ A}_{\Phi}|^{2}\right) \ , \\
S_{b}^{(v)} & = \frac{1}{2}E^{r}\bar{E}_{r}+\phi^{2}\frac{1}{2}|{ A}_{\Phi}|^{2}-\phi^{2}\frac{1}{2}\prescript{(3)}{}{A}^{r}\prescript{(3)}{}{\bar{A}}_{r} \ .
\end{align*}
Lastly we have the momentum and Hamiltonian constraint equations
\begin{align}
    \mathcal{H} &\coloneqq R-({\cal A}^2_a +2{\cal A}_b^2)+\frac{2}{3} K^2 -16\pi \rho =0 \ , \\
    \mathcal{M}_r &\coloneqq \partial_r {\cal A}_a -\frac{2}{3}\partial_r K +6{\cal A}_a\partial_r \chi + ({\cal A}_a - {\cal A}_b)(\frac{2}{r}+\frac{\partial_r b}{b})-8\pi j_r =0\ ,
\end{align}
where ${\cal A}_{ij}$ is the traceless part of the conformal extrinsic curvature with ${\cal A}_a ={\cal A}^r_r$ and ${\cal A}_b={\cal A}^\theta_\theta={\cal A}^\varphi_\varphi$.

\subsection{Numerical grid and stability}

For the numerical evolutions, we use the NADA 1D code, a code for numerical
relativity simulations in 1+1D in spherical symmetry that uses spherical coordinates described in
\citep{Montero:2012yr,SanchisGual:2014ewa,SanchisGual:2015sxa,SanchisGual:2015lje,EscorihuelaTomas:2017uac,DiGiovanni:2021vlu,DiGiovanni:2020frc}.
The BSSN and Klein-Gordon coupled equations are solved using a second-order
partially implicitly Runge-Kutta (PIRK) scheme \citep{CorderoCarrion:2012qac,CorderoCarrion2014}.
The evolutions are performed in a logarithmic grid, with a maximum
resolution of $\Delta r=0.05$, a time step of $\Delta t=0.3\Delta r$,
the number of radial points being $n_{r}=50000$ and the outer boundary
placed at $r_{{\rm max}}=10000$. The maximum time is $t_{{\rm max}}=10000$.
We impose radiative boundary conditions (Sommerfeld boundary conditions)
at the outer boundary \citep{Montero:2012yr,Alcubierre:2002kk}.

Regarding the perturbations applied to each model, we do
not apply any specific perturbation to the static models, since the
numerical truncation error is enough to break the staticity of the
models. However, later on, we will use forced perturbations by increasing
slightly the Proca fields in order to test the dynamical robustness
of certain configurations.

\section{Dynamical evolution and stability}
\label{sec4}

We proceed to evolve several models in different parts of the parameter
space in order to find out which configurations are stable. We will
evolve the models up to a maximum time $vt_{{\rm max}}=10000$. As our strategy, we shall first consider stars in the Proca limit, that is, when
$\alpha$ or $\lambda$ is very large, and then proceed to survey
the parameter space in the region where we have stars deviating from
the Proca limit. We fix $\omega=0.95$ in this analysis, which guarantees the examined stars always fall onto the candidate stable branch. We will plot the value of the Proca fields in natural
units ($c=\hbar=1$ and also with $G=1$), not dimensionless ones, in order to make
easier the comparison between models with different $\alpha$. This
is possible since we are effectively rescaling the gravitational constant
as in \cref{eq:new_einst}.

\subsection{Proca-Higgs stars in the Proca limit (negligible self-interactions)}

\begin{figure*}[tpbh]
\begin{centering}
\input{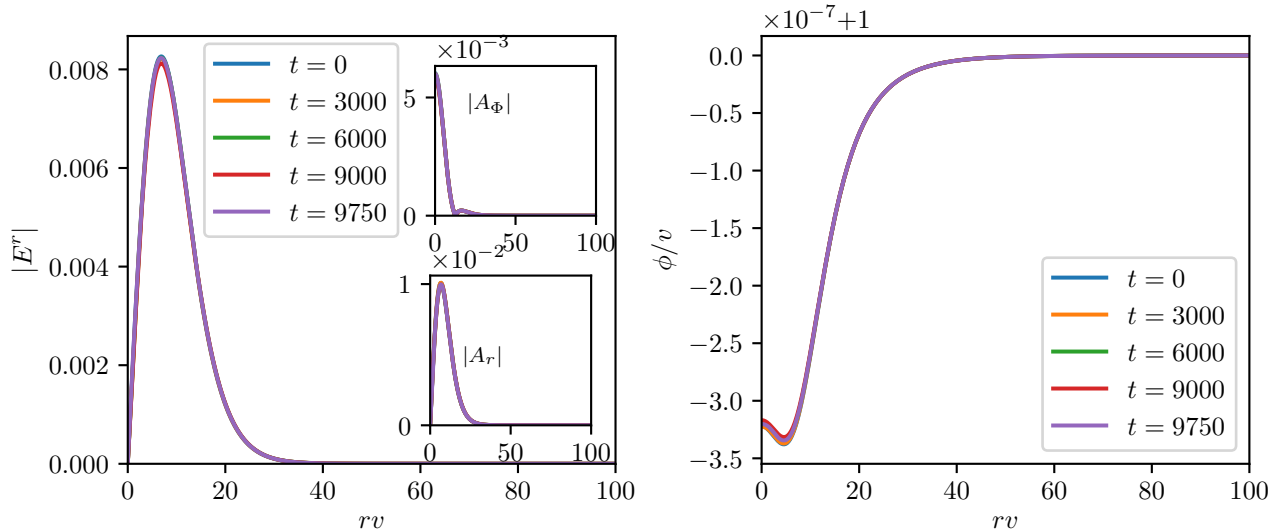}
\par\end{centering}
\caption{\label{fig:alpha=300-lambda=0.005-omega=0.95}Proca
and Higgs fields for $\alpha=300$, $\lambda=0.005$, $\omega=0.95$, for various times of the evolution.}
\end{figure*}

\begin{figure*}[tpbh]
\begin{centering}
\input{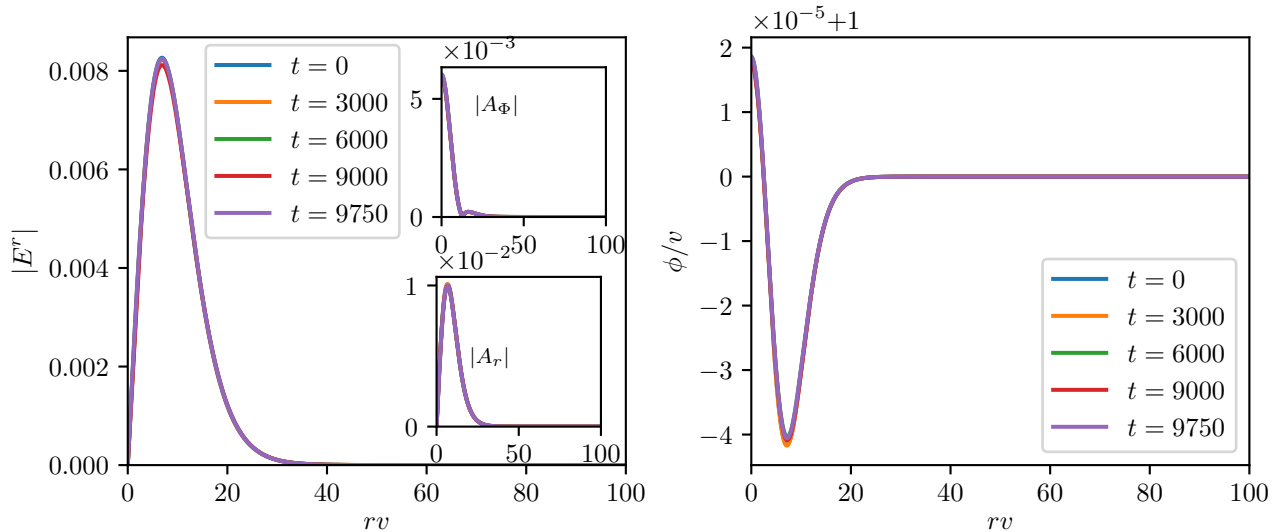}
\par\end{centering}
\caption{\label{fig:alpha=0.35-lambda=100-omega=0.95}Proca
and Higgs fields for $\alpha=0.35$, $\lambda=100$, $\omega=0.95$, for various times of the evolution.}
\end{figure*}

As the VEV increases our solutions approach a mini-Proca star, so
it is important to confirm that we obtain a stable model against radial
perturbations. Moreover, at low energies ($\alpha\to\infty$ or $\lambda\to\infty$)
we recover approximately the free Proca model with mass $v$, since
the self-interactions will be negligible compared to the mass term
(approximately six orders of magnitude smaller for the $\alpha=300$ model and four
orders for the $\lambda=100$ one).
We can see in \cref{fig:alpha=300-lambda=0.005-omega=0.95,fig:alpha=0.35-lambda=100-omega=0.95}
the time evolution of the radial part of the Proca-Higgs electric and scalar fields when $\alpha$ or $\lambda$
are very large, respectively.

In both cases, the scalar field is frozen at the VEV, determining the effective
mass of the vector boson and the stars are stable (show no evolution), as expected, since
they reduce to mini-Proca stars in this limit, see \cref{fig:Mass-vs-frequency}.
It is important to emphasize that both solutions agree with each other,
having equal magnitudes for the Proca fields and, although the scalar
field deviates more from the VEV in the very large $\lambda$ case,
the difference between the scalar field and the VEV is negligible,
thus each model gives approximately the same metric functions. Our
model is consistent with what we would expect from the pure Proca
model with a mass of $\mu=\alpha/\sqrt{4\pi G}$, which is stable under
radial perturbations. Adding forced perturbations to this model does not change the final state; however, strong perturbations (more than 3\% of the initial values of the fields) do not allow the star to relax completely in the time period considered for the evolution.

\subsection{Proca-Higgs stars with relevant self-interactions}

For smaller values of $\lambda$ and $\alpha$, we move away
from the Proca limit, getting stars with non-negligible self-interactions. One interesting question is whether stars in
this region of the parameter space are stable, and if not, identify
the regions of the parameter space where such is the case. The domain
of existence of Proca-Higgs stars can be seen in \cref{fig:Mass-vs-frequency}.

\subsubsection{Stable stars}

\begin{figure*}[tpbh]
\begin{centering}
\input{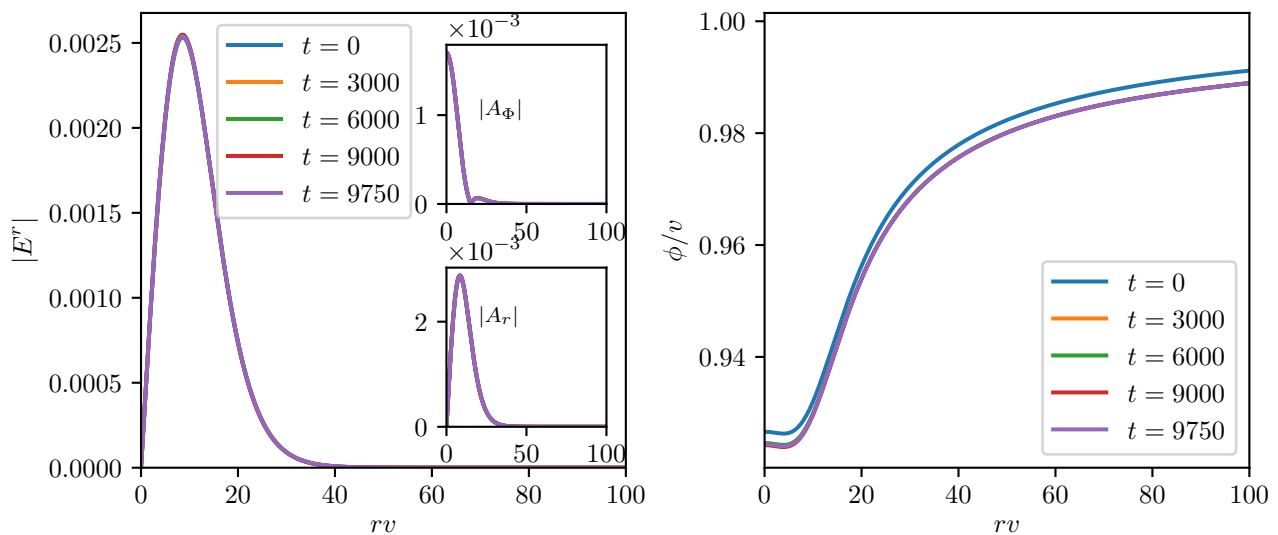}
\par\end{centering}
\caption{\label{fig:alpha=0.35-lambda=0-omega=0.95}Proca
and Higgs fields for $\alpha=0.35$, $\lambda=0$, $\omega=0.95$, for various times of the evolution.}
\end{figure*}

Even in this regime, we can find stable stars. Consider first
a boson star with parameters $\alpha=0.35$, $\lambda=0$, $\omega=0.95$.
This is just the case of a complex Proca field coupled with a free
scalar field. This model lies on the candidate stable branch, and it
is indeed stable as one can see in \cref{fig:alpha=0.35-lambda=0-omega=0.95}.
However, the scalar field is not frozen to the VEV anymore. Since we
lose the scalar potential, the vector field cannot acquire mass dynamically.

\begin{figure*}[tpbh]
\begin{centering}
\input{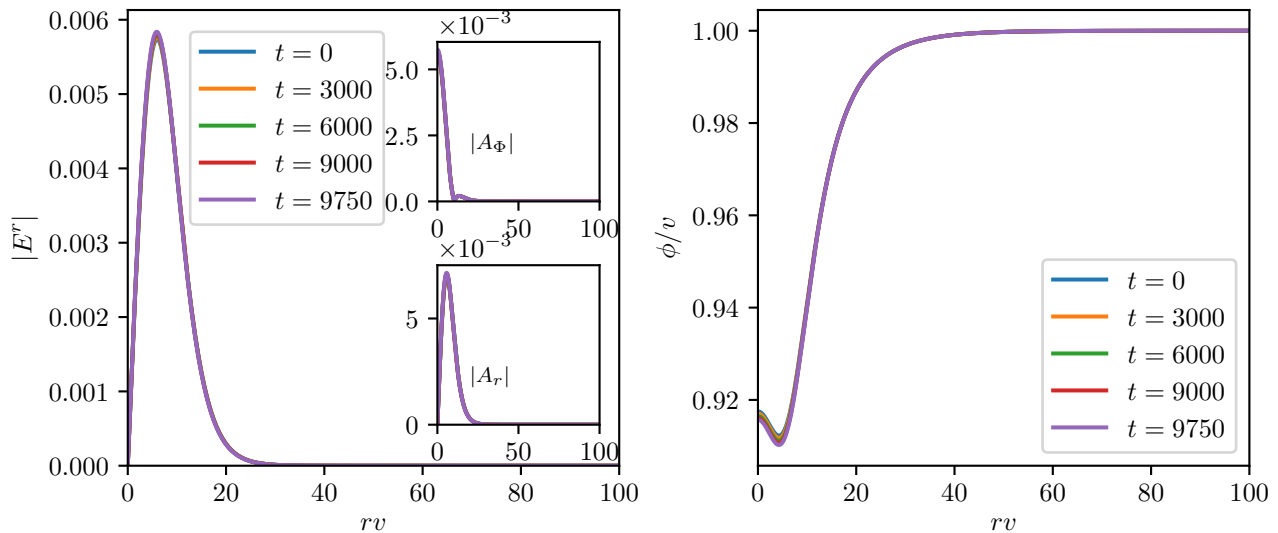}
\par\end{centering}
\caption{\label{fig:alpha=0.35-lambda=0.005-omega=0.95}Proca
and Higgs fields for $\alpha=0.35$, $\lambda=0.005$, $\omega=0.95$, for various times of the evolution.}
\end{figure*}

We can also see the model with $\alpha=0.35$, $\lambda=0.005$, $\omega=0.95$,
which is more relevant since the scalar field goes to the VEV at infinity.
This model is also stable with the scalar field going to $v$ as the
distance to the center increases. However, at the center it takes values
different from the VEV - see \cref{fig:alpha=0.35-lambda=0.005-omega=0.95}.
In fact for values of $\lambda$ close to zero all the Proca-Higgs
stars studied are stable. Near zero the effective theory breaks down since
the self-interactions are much stronger. For the model at hand the
self-interaction term is about half the mass term.

\subsubsection{A collapsing model}

\begin{figure*}[tpbh]
\begin{centering}
\input{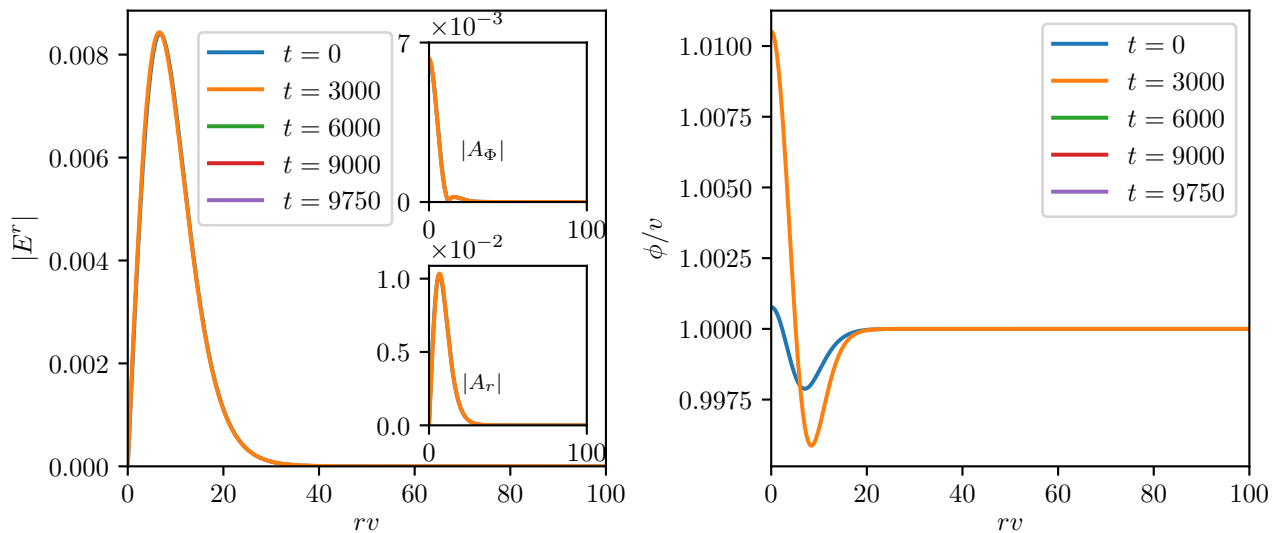}
\par\end{centering}
\caption{\label{fig:alpha=0.35-lambda=2-omega=0.95}Proca
and Higgs fields for $\alpha=0.35$, $\lambda=2$, $\omega=0.95$, for various times of the evolution.}
\end{figure*}

\begin{figure*}[tpbh]
\begin{centering}
\input{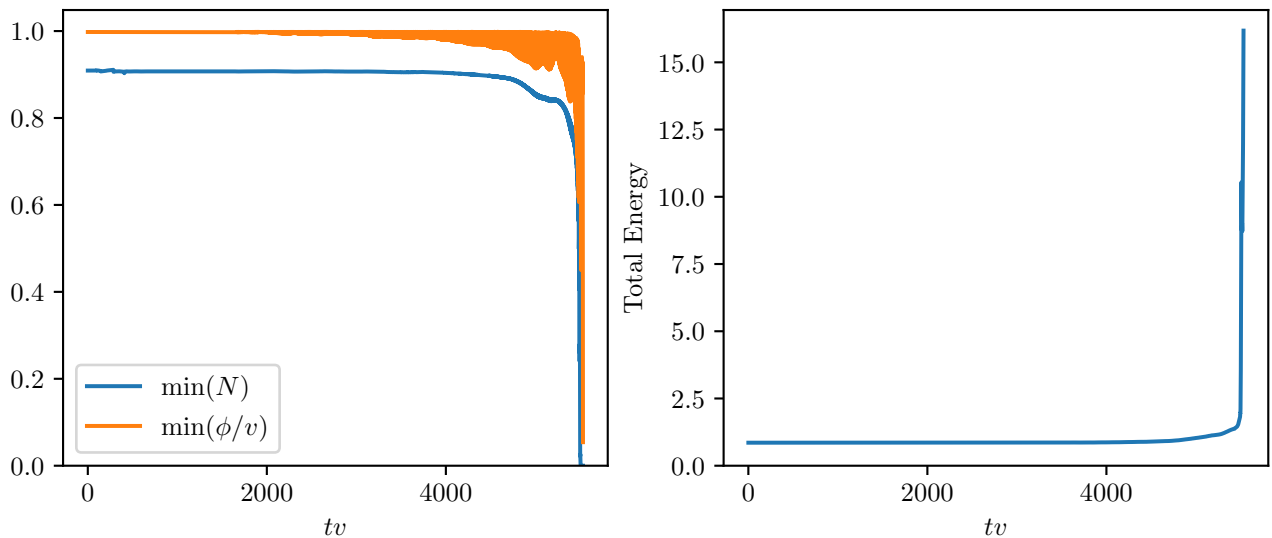}
\par\end{centering}
\caption{\label{fig:Minimum-value-of-sphi}Minimum value of the scalar field
and of the lapse (left). Integrated energy density over the whole spacetime (right).}
\end{figure*}

The model with $\alpha=0.35$, $\lambda=2$, $\omega=0.95$ is unstable. We find evidence of collapse into a black hole, even though our formulation does not allow us to follow the evolution past the event horizon formation -- see \cref{fig:alpha=0.35-lambda=2-omega=0.95}.
This inability (leading to the code crashing) is triggered by the emergence of points where the
scalar field $\phi$ vanishes, as seen in \cref{fig:Minimum-value-of-sphi}.
Since in our equations of motion we have explicit terms of $1/\phi$, coming from \cref{eq:one-over-sphi},
the equations blow up. However, just before the crash we have all the
telltale signs of the formation of an apparent horizon, namely the minimum
value of the lapse going to zero,
the total energy density starting to diverge (\cref{fig:Minimum-value-of-sphi})
at the origin, where a large density region is emerging, and  the position of the apparent horizon at the moment of the
crash can be computed. Indeed, from \cref{fig:Minimum-value-of-sphi}
we see that the lapse reaches zero instants before the scalar field,
showing that it is the collapse of the star that drives the scalar
field to zero and not vice-versa; that is, it is not the vanishing
scalar field that triggers a collapse.
The post collapse evolution is, however, inaccessible to our current
formulation of the problem. We emphasise that this instability
occurs on the \textit{candidate stable branch} for $\lambda=2$ as
one can see in \cref{fig:Mass-vs-frequency}. This is unexpected since
for smaller values of $\lambda$ we have stable boson stars, and we
also know that in the large $\lambda$ limit, the stars approach pure
Proca stars. This hints that there might be a region of unstable stars in the parameter space and also shows that these models
are not stable in the whole parameter space.

\subsubsection{Regions of instability}

The existence of an unstable model ($\alpha=0.35$, $\lambda=2$, $\omega=0.95$)
prompts us to restrict our search in the nearby regions of the parameter
space, since we know that for very small and very large $\lambda$
we obtain stable solutions. This might hint the existence of an instability
region or instability bands. For models in the region of interest the
self-interactions are at approximately two orders of magnitude smaller than the mass term.

We proceed with a survey of points in
the parameter space close to $\alpha=0.35$ and $\lambda=2$ and evolve
models varying $\alpha\in[0.2,0.5]$ in steps of $0.05$, varying
$\lambda\in[1.0,3.0]$ in steps of $0.1$ and $\omega=0.95$. Our first conclusion  -- 
\cref{fig:Instability-island-w/o-perturbation} -- is that whereas some models collapse in a similar manner as the example described above and some others show visible changes but without clear evidence of collapse, there are still models that show no evolution. To further test the latter, since the numerical truncation
error was not enough to trigger a visible change in their evolution, we chose to add a perturbation by hand.
The perturbation was enforced in
the Proca field by increasing the $f$ and $g$ fields (see \cref{eq:proca-decomp-initial})
by one, two and three percent. Such extra perturbation indeed triggered instabilities in otherwise unchanged stars -- 
\cref{fig:Instability-island-w/o-perturbation}.
The legend in \cref{fig:Instability-island-w/o-perturbation} refers to the final states of the evolution and if a percentage is shown it means a forced perturbation was applied to the initial data of the model where the percentage is the relative strength of the perturbation.
Still, even with the forced perturbations
some stars remain around the static configuration or are, despite some visible changes, in
a non-equilibrium state by the end of the evolution, without a clear hint to what their final state may be. 

A decision on the endpoint of the latter evolutions, and a clearer verdict on the islands of stars that remain stable even with the forced perturbations in \cref{fig:Instability-island-w/o-perturbation} can, in principle, be made by increasing the duration of the evolutions. But the main point we wish to make -- that we have solutions
that are unstable in the candidate stable branch, in this region of the parameter space where self-interactions are very important -- is crystal clear. Moreover, we can observe that the evolution of the models in this region of the parameter space is very
sensitive to initial conditions, a feature also known in other bosonic star models.

\begin{sidewaysfigure*}
\begin{centering}
\input{instab_island.tex}
\end{centering}
\caption{\label{fig:Instability-island-w/o-perturbation}Instability regions
without and with forced perturbations.}
\end{sidewaysfigure*}

\section{Discussion and Conclusion}
\label{sec5}

In this paper, we have embarked on a test of the healing power of self-interactions in bosonic star models. A  pattern has been observed for scalar boson stars that such self-interactions mitigate dynamical instabilities making the models more robust~\cite{Siemonsen:2020hcg,Sanchis-Gual:2021phr,Ildefonso:2023qty}. Here we considered vector boson stars, aka Proca stars. Self-interactions can be consistently introduced using the Einstein-Proca-Higgs model~\cite{Herdeiro:2023lze} that avoids  the problems in Einstein-Proca models where the Proca field has self-interactions~\cite{Coates:2022nif,Clough:2022ygm,Mou:2022hqb,Coates:2022qia,Barausse:2022rvg}.

We have seen that there are regions in the parameter space $(\lambda,\alpha)$
where we have unstable stars in the naive candidate stable branch, notwithstanding the fact that for very
small and for very large $\lambda$ we have stable Proca-Higgs stars. We cannot point out a definite reason for this behavior, but it might be similar to the instability bands observed in
\citep{Sanchis-Gual:2021phr,Nambo:2024gvs}.

The identified regions of instability are very sensitive to initial conditions. We have verified that  when we apply a forced perturbation we can see  previously stable
models collapsing into a black  hole or becoming unstable. 

Despite a very sensitive picture to initial conditions, our analysis corroborates that self-interactions have the potential to destabilize Proca stars,
since as we increase the self-interactions, the stars, which are stable in the
mini-Proca limit, become unstable when $\lambda\sim 1$. But observe they regain stability when $\lambda\to0$. This is unlike scalar
excited boson stars where self-interactions always stabilize those
configurations after a certain self-interaction strength.

From another point of view, near the Proca limit, these stars are stable and reduce to Proca,
showing its usefulness as a UV completion of the Proca theory.

Let us close by remarking that it was recently observed that if one allows \textit{non-spherical} dynamics even spherical mini-Proca stars are not stable, decaying to a non-spherical (prolate) configuration~\citep{Herdeiro:2023wqf}. For the Proca-Higgs stars, this means that even for the configurations we have seen as stable under spherical dynamics, there may be a non-spherical decay channel. Whereas this remains an interesting problem for further study, it does not contradict our main message here, which is that the impact of self-interactions is critically different in the scalar and vector cases of bosonic stars.

\begin{acknowledgments}
This work is supported by the Center for Research and Development in Mathematics and Applications (CIDMA) through the Portuguese Foundation for Science and Technology (FCT -- Fundaç\~ao para a Ci\^encia e a Tecnologia) through projects: UIDB/04106/2020 (with DOI identifier \url{https://doi.org/10.54499/UIDB/04106/2020}); UIDP/04106/2020 (DOI identifier \url{https://doi.org/10.54499/UIDP/04106/2020});  PTDC/FIS-AST/3041/2020 (DOI identifier \url{http://doi.org/10.54499/PTDC/FIS-AST/3041/2020}); CERN/FIS-PAR/0024/2021 (DOI identifier \url{http://doi.org/10.54499/CERN/FIS-PAR/0024/2021}); and 2022.04560.PTDC (DOI identifier \url{https://doi.org/10.54499/2022.04560.PTDC}). This work has further been supported by the European Horizon Europe staff exchange (SE) programme HORIZON-MSCA-2021-SE-01 Grant No.\ NewFunFiCO-101086251.
M.B.\ is supported by the FCT grant 2022.09704.BD. N.S.G. acknowledges support from the Spanish Ministry of Science and Innovation via the Ram\'on y Cajal programme (grant RYC2022-037424-I), funded by MCIN/AEI/10.13039/501100011033 and by ``ESF Investing in your future”. N.S.G. is further supported by the Spanish Agencia Estatal de Investigaci\'on (Grant PID2021-125485NB-C21) funded by MCIN/AEI/10.13039/501100011033 and ERDF A way of making Europe.
M.Z.\ acknowledges financial support through FCT grant 2022.00721.CEECIND with DOI identifier \url{https://doi.org/10.54499/2022.00721.CEECIND/CP1720/CT0001}. E.S.C.F.\ is supported by the FCT grant PRT/BD/153349/2021 under the IDPASC Doctoral Program. Computations have been performed at the Argus
cluster at the U.~Aveiro and at the Navigator Cluster at the LCA in U.~Coimbra through projects 2021.09676.CPCA and 2022.15804.CPCA.A2.
\end{acknowledgments}

\appendix

\section{Numerical Convergence}

\begin{figure}[tpbh]
\begin{centering}
\input{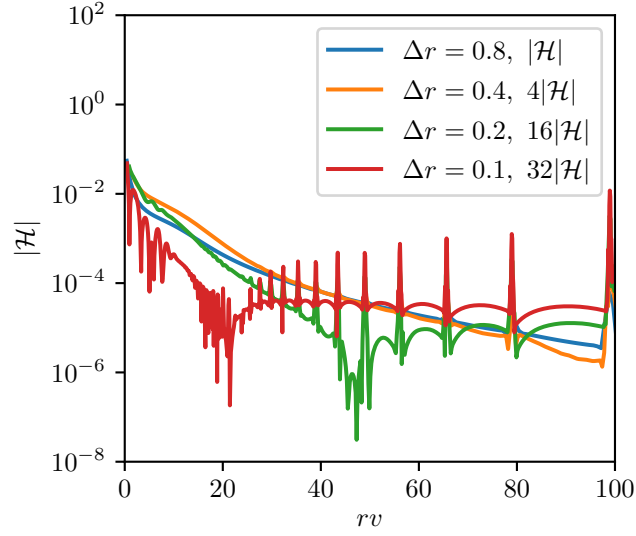}
\par\end{centering}
\caption{\label{fig:Hamiltonian-constraint-convergen}Hamiltonian constraint
convergence at $t=1200v$.}
\end{figure}

To assess the quality of the numerical simulations we perform a convergence
test consisting of comparing different quantities from various grid
resolutions and checking if the results converge to the expected value following the convergence order given by the spatial derivatives and the time integration.
In order to perform our numerical evolutions we imported initial data
into the code which was then interpolated to the evolution grid. We
consider only numerical error coming from the finite difference operations,
which dominates the error if we use resolutions coarser than the initial
data resolution. In \cref{fig:Hamiltonian-constraint-convergen} we
show the absolute value of the Hamiltonian constraint for four different
resolutions at the instant $t=1200v$ for the model $\alpha=300$, $\lambda=0.005$, $\omega=0.95$
which is practically a mini-Proca star, and we find a second order
convergence for all resolutions, except at $\Delta r=0.1$ for which we get
between first and second order because for high resolutions we are
not improving anything compared with the initial grid, which we are keeping fixed.

\begin{figure}[h!]
\begin{centering}
\input{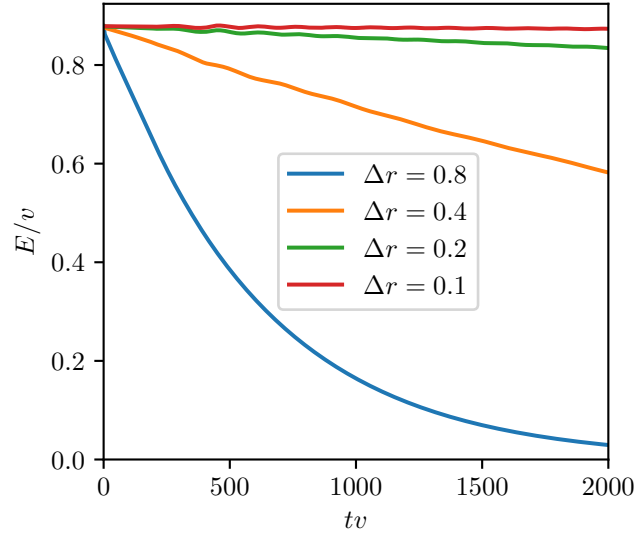}
\par\end{centering}
\caption{\label{fig:Evolution-of-the}Evolution of the integrated field energy
density for different resolutions.}
\end{figure}

\begin{figure*}[tpbh]
\begin{centering}
\input{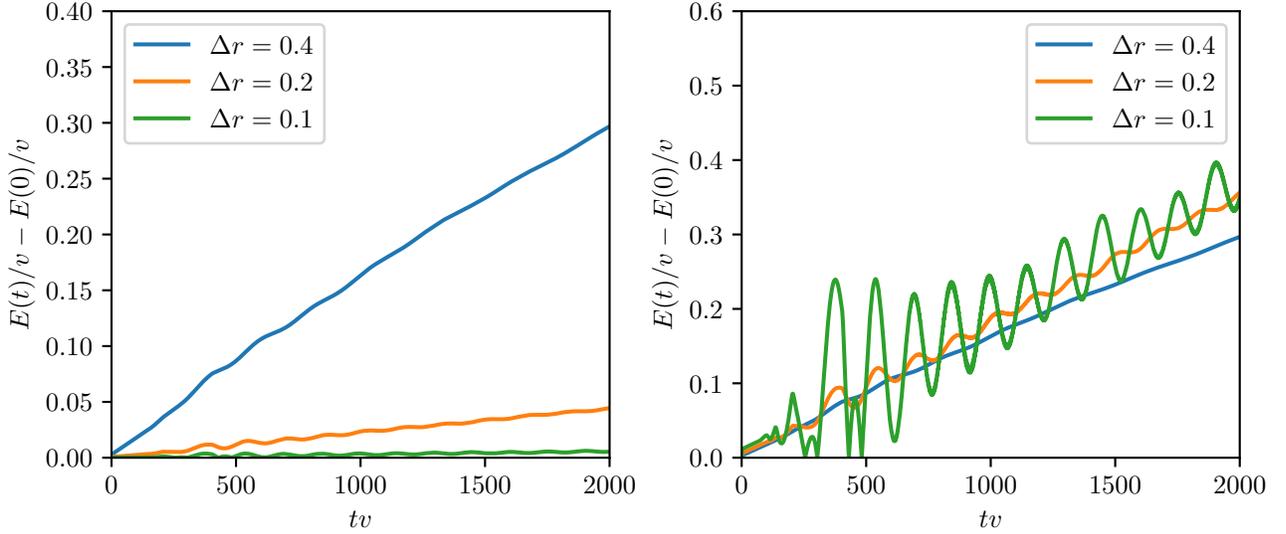}
\par\end{centering}
\caption{\label{fig:Difference-between-total}Difference between total energy
and total initial energy (left) and re-scaled functions to third-order
convergence (right).}
\end{figure*}

We compare also the drift of the conservation of the field energy as
in \cref{fig:Evolution-of-the}. However, due to numerical error, the
mass decreases with time, and for low resolutions the numerical solutions
are not good enough as the mass decreases too fast. Taking the deviation with respect to the initial value of
the energy $E(t=0)$, we find that the order of convergence is 3, due
to the fourth-order interpolation, the second-order PIRK and the fourth-order finite differencing. The scaled functions can be seen in \cref{fig:Difference-between-total}. Similar results were found in~\cite{Brito:2023fwr}.

\bibliographystyle{aipnum4-2}
\bibliography{refs}

\end{document}